\documentclass[iop]{emulateapj}

\begin{document}

\title{ Warm Absorber Diagnostics of AGN Dynamics}

\author{T. Kallman, A. Dorodnitsyn \altaffilmark{1}}

\altaffiltext{1}{NASA/GSFC, Code 662, Greenbelt MD 20771}


\begin{abstract}
Warm absorbers and related phenomena are one of the observable 
manifestations of outflows or winds from active galactic nuclei (AGN).
Warm absorbers are common 
in low luminosity AGN, they have been extensively studied observationally, 
and are well described by simple phenomenological models.  However, major 
open questions remain:  What is the driving mechanism?  What is the 
density and geometrical distribution?  
How much associated fully ionized gas is there?  What is the relation 
to the quasi-relativistic 'ultrafast outflows' (UFOs)?  
In this paper we present  synthetic spectra for the 
observable properties of warm absorber flows 
and associated quantities.   We use 
ab initio dynamical models, i.e. solutions of the equations of motion for gas
in finite difference form.  The models employ various plausible assumptions
for the origin of the warm absorber gas and the physical mechanisms affecting 
its motion.  The synthetic spectra are presented as an observational test 
of these models.  In this way we explore various scenarios
for warm absorber dynamics.   We show that observed spectra place certain requirements 
on the geometrical distribution of the warm absorber gas, and that not all 
dynamical scenarios are equally successful at producing spectra similar to what 
is observed. 
\end{abstract}

\section{Introduction}

Warm absorber spectra in the X-ray band consist of absorption features 
from gas with a range of ionization states, from near-neutral to 
hydrogenic for medium-Z elements, with line widths and blueshifts $\sim$1000 km s$^{-1}$.
They are seen in a significant fraction of the AGNs which have been observed with 
sufficient sensitivity to detect them \citep{Reyn97, Ricc17, blus05, Mcke07}.  

The blueshift implies outflow, and the existence 
of outflows from objects thought to be powered by accretion raises questions about the 
outflow origin, how it interacts with the accretion flow, and whether there is a causal 
connection between the two flows.
Theoretical understanding of warm absorbers is complicated by the
absence of spectral diagnostics which uniquely constrain the location of the absorber 
relative to the black hole.  Information about the conditions in the flow can be derived using
the assumption that the gas is in ionization equilibrium and 
the ionization and heating are supplied by the strong continuum from the central black hole.  
If so, the phenomenological fits to the spectra 
constrain the ionization parameter of the gas, i.e. the ratio of the 
gas density and dilution factor of the illuminating radiation.
This plus the observed speed allows the mass flux of the flow to be estimated.
The location can also be estimated using the line widths 
and blueshifts, assuming they are approximately virial.  This implies a location 
near $\sim$ 1 pc for a 10$^6$ M$_\odot$ black hole.   Variability of the line 
strengths or their ratios can also be used to estimate the warm absorber location 
for some objects, and predicts values which can differ significantly from the virial 
estimates \citep{kron07, beha03}.

Mechanisms which can account for the driving of warm absorber outflows include
magneto-rotational models \citep{Fuku10} in which the flow 
is tied to a strong, rotating, externally imposed magnetic field and is 
driven outward by centrifugal forces; thermal \citep{Bals93} or 
thermally driven models in which the flow is evaporated by X-ray heating \citep{Bege83}; and radiation 
pressure driven winds \citep{Murr95,Prog00,Prog02}, in which the flow is 
driven primarily by radiation pressure in lines.
These flows may originate from a geometrically thin accretion disk, and/or may be 
associated with the obscuring $\sim$1 pc torus either by direct X-ray heating 
or by radiation pressure on trapped infrared (IR) \citep{Elit06,Krolik07}.
Warm absorber gas may also be associated with entrainment by higher 
velocity gas \citep{Risa10}, which is not as easily observed.

The obscuring torus location is constrained by interferometric observations in some objects 
\citep{Jaff04}.  It is approximately coincident  with the dust sublimation radius, and so 
can produce both the observed line widths and the strong dynamical effects associated 
with IR radiation pressure on dust.

We have constructed  hydrodynamic models for warm absorber gas using 
assumptions based on these current ideas:  That the warm absorber gas 
represents an ionized component of the gas responsible for 
AGN unification and obscuration, with the ionization supplied by X-ray illumination from close to the black hole 
horizon, and that the outflow arises from thermal evaporation supplemented by radiation pressure 
\citep{ Dorodnitsyn08a, Dorodnitsyn08b, Dora08_3, 
DoraKall09, DoraKall10, DoraKall16, DoraKall17}.   
These models incorporate the physics of X-ray heating and radiative  cooling, radiative driving, 
toroidal geometry, and attenuation of the radiation from the central black hole.
This is combined with an ab initio treatment of the hydrodynamic or magnetohydrodynamic (MHD) equations of motion in 
2.5 or three dimensions using either the {\sc zeus} or {\sc athena} code.
 Attenuation of the illuminating 
radiation  from the central black hole is treated using a simple 
single stream solution of the radiative transfer equation.

A key check on our models, and others, is to what extent they can account for 
the details of the observed spectra.
We carried out such a check for our early models, which were pure hydrodynamics 
in 2.5 dimensions \citep{Dora08_3}.  More recently, we have constructed MHD models, and
in this paper we present calculations of the X-ray spectra which would be produced by 
these models.  We use the output of the dynamical models,
i.e. the density and velocity fields, and do a three-dimensional ray trace calculation of 
the spectrum seen by a distant observer from various viewing angles.  In Section \ref{observations} 
we summarize the aspects of warm absorber spectra observations 
which are relevant for our later comparisons.  
In Section \ref{models} we summarize the 
MHD/Hydro models we use for the spectrum calculations.  In Section \ref{calculations} we describe
our spectrum synthesis calculations.  In Section \ref{results} we describe the results.

\section{Observable Properties}\label{observations}

An example warm absorber spectrum is  the Seyfert 1 galaxy NGC 3783
\citep{ Kasp02} taken with the 
the $Chandra$ high energy grating (HETG).  
This spectrum contains $\geq$100 
easily identifiable X-ray absorption features from
the K shells of highly ionized ions of medium-Z elements:
O - S, and also from the L shell of 
partially ionized iron.  

A fit to this spectrum consists of
with three components of gas in photoionization equilibrium, all outflowing 
from the  AGN at $\simeq$600 km s$^{-1}$. 
This model produces general agreement with the data, for the continuum and the strong lines, 
although it is  only marginally statistically acceptable ($\chi^2/\nu\simeq$ 1.6).
Photoionization models are parameterized in terms 
of the ionization parameter $\xi=4 \pi F/n$, where $F$ is the X-ray energy flux between 1 and 1000 Ry and $n$ 
is the gas number density (Tarter Tucker and Salpeter 1968), and the 
high ionization components in the fit to NGC3783 has  ${\rm log}(\xi) \simeq 2$ and 2.5.
The low ionization state component has ${\rm log}(\xi) \simeq 1$. 
The observations strongly rule out the existence of gas at intermediate ionization parameters, i.e ${\rm log}(\xi) \sim$1.5.
Similar results have been found in spectra from other objects as well \cite{Mcke07, Holc07}.   


 X-ray observations 
of the Seyfert 2 galaxy NGC 1068 show spectra which are consistent with viewing the same gas 
seen in Seyfert 1 galaxies from a different angle, 
i.e. with the continuum source blocked so that the spectrum is dominated 
by emission \citep{Kink02}.  Fits of multiple-component models to 
$Chandra$ spectra from NGC 1068 imply similar properties for the gas in Seyfert 1s and Seyfert 2s,
with the feature that Seyfert 2s appear to require a range of gas densities coexisting in a given 
spatial region \citep{Ogle03, kall14}.


The model fits discussed so far demonstrate the existence of a 
range of ionization states in warm absorber gas, and that photoionization equilibrium 
models fit many of the lines.
The results of these fits, however, provide only indirect information about the flow itself.
The column densities,  ionization parameters, and turbulent and outflow velocities
are all treated as free parameters and are adjusted to obtain the 
best fit.  Their values can be tested against the predictions of 
dynamical models, but they are not derived directly from physical models for the 
origin of this gas. 


\section{Dynamical Models}\label{models}

The dynamical models used here employ the same tools as used in \cite{DoraKall17}. 
We investigate the evolution of a magnetized torus exposed
to external X-ray illumination via three-dimensional numerical
simulations.  To solve a system of equations of ideal
magnetohydrodynamics
we adopt the
second-order Godunov code {\tt Athena} version 2\citep{Ston08}.  The code
is configured to adopt a uniform cylindrical grid,
$\{R_{i},\thinspace\phi_{j},\thinspace
z_{k}\}$,\citep{Skin10}.

Simulations are performed in a static Newtonian gravitational
potential.  The van Leer integrator is found to be the most robust
choice for the  simulations presented in this work.  We
implement the heating and cooling term in the same way as the
static gravitational potential source term is implemented in the
current van Leer integrator implementation in {\tt Athena}. That is, the heating source term
is implemented to guarantee 2-nd order accuracy in both space and time.


In our simulations the inner, illuminated surface of the 
torus forms a funnel which efficiently launches and collimates an outflow.
Gas which flows close to the axis acquires a high terminal speed, 
$\sim 1000 \,\rm km\, s^{-1}$, but is too highly ionized to produce 
observable warm absorber spectra.  Gas which is viewed at larger angles 
with respect to the axis (greater inclinations) can produce 
warm absorbers.
This is due to the combined effects of the greater density
(and hence lower equilibrium ionization state), self-shielding of the flow, 
and departures from thermal equilibrium in the flow.  When viewed at high inclinations,
i.e. close to 90 degrees, the line of sight passes through gas which has a 
Thomson depth $\sim$10.  This gas has low ionization and is effectively opaque 
to all continuum from the central source.

In our work so far we have modified Athena to include X-ray heating, radiative cooling and radiation 
pressure forces.  In doing so, we assume that the gas 
is locally in ionization equilibrium, but not necessarily in thermal equilibrium. 
This is justified by the fact that the slowest thermal processes involve Compton 
scattering, and these are independent of the gas ionization state.
That is, the gas energy equation includes the effects of 
radiative heating and cooling, along with adiabatic and advection effects.

A key goal of our work has been to understand the 
warm absorber flow, and its implications for the AGN accretion flow 
and mass budget.   The starting point for our models is the simple hypothesis that the 
warm absorber flow represents gas which is evaporated from a relatively cold, geometrically and optically thick
torus.  That is, the warm absorber flow itself is driven primarily by thermal evaporation due to 
X-ray heating, but that radiation pressure and the effects of magnetic fields also affect the dynamics.
Our models confirm this crudely:  the warm absorber gas occupies much of the `throat' of the torus, 
i.e. the region outside the cold torus, closer to the axis.  The throat
$\sim$half the solid angle seen from the central objects.  The warm absorber gas has ionization parameter in the 
range -1$\leq {\rm log}(\xi) \leq$4; the density is highest and the ionization parameter is lowest closest to the 
cold torus surface.  Lines of sight which traverse this region produce warm absorber spectra most nearly 
resembling observations \citep{Dorodnitsyn08a}.  The outflow speeds in the throat are comparable to the sound 
speed at temperatures $10^5$ -- $10^7$K, i.e. 100 -- 1000 km s$^{-1}$.

However, in addition we have found that the character of the warm absorber is
closely coupled with the structure of the torus.  This is because  
the existence and character of the flow  depends on the shape 
of the torus via the  divergence of the flow streamlines, 
the strength and incident angle of the X-ray illumination, 
and on the strength and direction of the effective gravity.  
At the same time, the flow can carry away a significant amount of mass, and so will 
deplete the torus outer layers, thereby affecting the shape of the torus.
So the torus interior cannot be considered as a boundary condition 
(e.g. as in \citet{Bals93}); rather, we need to include it in the 
computational domain.  
We have done this 
by using as our initial condition the constant angular momentum adiabatic structure 
suggested by \citet{Papa84} and adopted by  \citet{Ston99} in the context of non-radiative 
flows. 

Our models do not employ any hydrodynamic viscosity.  The MHD formulation produces a robust 
magnetorotational instability (MRI) which leads to inflow from dense regions near the midplane 
of the distribution, while evaporative flow occurs from moderate and high latitudes.  Details of
these dynamics are described in detail in \citet{DoraKall17}.  Models are evolved for $\sim$60-80 
dynamical times, where the dynamical time is given by $t_{dyn}=\sqrt{R^3/2GM}=10^{11} R_{pc}^{3/2} M_7^{1/2}  {\rm s}$ and
$R_{pc}$ is the fiducial radius in pc and $M_7$ is the mass of the black hole in $10^7 M_\odot$.
We find that the time evolution of our models is sensitive to the initial configuration of the magnetic 
field.  We consider two different choices: one in which the poloidal field strength is proportional to the gas 
density in order to produce an approximately constant ratio of magnetic pressure to gas pressure (called TOR), 
and one in which the field is an approximate solution to the Grad-Shafranov equation as implemented by 
Soloviev (called SOL; see \citet{DoraKall17} for details).  
In the latter case, the field configuration evolves much more slowly and results in 
a more compact and long-lived torus.    We emphasize that 
the choice of field geometry is only an initial condition; after the simulation is begun the field evolves 
according to the MHD solution prediction.  Here and in what follows we explore these, and also compare with model 
in which the MHD treatment is turned off, which we refer to as a pure hydrodynamic model.  The total mass of gas in the initial torus is the same for all the models of a given type, i.e.   SOL, TOR or hydro.

\section{Spectrum Synthesis}\label{calculations}

Synthesizing the spectrum provides a key test for dynamical models.  {\sc xstar}
provides a full frequency dependent opacity 
and emissivity  calculation which is coupled to the ionization and thermal balance  
as a standard component of the calculation of the gas ionization balance.  
A dynamical calculation predicts the velocity field and the 
spatial distribution of the gas density.  We use this to calculate
the emissivity and opacity of the gas in both continuum 
and lines as a function of position, and we use this to calculate 
synthetic spectra by integrating over 
the wind using the formal solution to the equation of transfer.  
Scattered emission from lines is 
calculated using the Sobolev approximation source function described by \citet{Cast70}.   
This procedure allows a direct test 
of the dynamical scenario without the intermediary step of parameterizing the 
properties of the flow, and then using the parameterized velocity and density field
to calculate the spectrum.    Our calculations include emission and absorption due to 
both resonance scattering in lines and thermal emission and absorption associated with electron impact 
excitation and bound-free transitions. We can test for the importance 
of emission filling in of absorption lines, and also the appearance of the flow 
from various inclinations.  

We calculate the specific luminosity seen by a distant observer by solving the 
 formal solution to the equation of transfer.

\begin{equation}
\label{eq1}
L(\varepsilon) 
= \int dV \kappa(\varepsilon,{\bf r}) S(\varepsilon,{\bf r}) e^{-\tau(\varepsilon,{\bf r})}
\end{equation}

\noindent where $\varepsilon$ is the photon energy, $S(\varepsilon,{\bf r})$,  is the source function, $\kappa(\varepsilon,{\bf r})$
is the opacity,  $\tau(\varepsilon,{\bf r})=\int{\kappa(\varepsilon,{\bf r}) d\zeta}$ where $\zeta$ is the 
physical distance along the line of sight, is the optical
depth from a point ${\bf r}$ to a distant viewer,
The luminosity seen by a distant observer is the total energy (in ergs s$^{-1}$ sr$^{-1}$) radiated by the
system in that direction.  Equation \ref{eq1} defines the scattered luminosity, $L$
(polarized plus unpolarized); observations are also affected by an
unscattered component, $L_u(\varepsilon)=L_0(\varepsilon)e^{-\tau(\varepsilon,{\bf r_x})}$
where ${\bf r_x}$ is the position of the X-ray source. 

The source function is calculated by integrating outward from the center and calculating the attenuation: 

\begin{equation}
\label{eq2}
S(\varepsilon, {\bf r}) = L_0/(4\pi r^2) e^{-\tau_0(\varepsilon,{\bf r})}
\end{equation}

\noindent where $L_0$ is the specific luminosity emitted by the central source, 
and $\tau_0(\varepsilon,{\bf r})$ is the optical depth from the center along a radial ray.
This formulation assumes at most a single scattering of the primary X-rays as they 
traverse the gas.  This assumption is not valid if the the optical depth from the X-ray source is
very large.  In that case, accurate treatment of the multiply scattered radiation requires use of 
a different computational technique, eg. Monte Carlo \citep{higg18}.  Since the gas in our simulations 
does have Thomson depth $\sim$10 when traversed in the orbital plane, our technique does not accurately 
treat this process.  However, the gas in the high optical depth region of our models is neutral, and scattering 
is only comparable to absorption at energies $\geq$ 6 keV.   
In what follows we do present results for 
spectra from models viewed at high inclination; in this case most of the flux seen by a distant observer 
comes from radiation which scatters once, 
or is absorbed and reemitted once.  The primary site for this is the torus throat, as defined in section 
\ref{models}. As we will show, the torus throat has a range of ionization and density conditions.  
Gas closest to the cold torus, which is traversed by lines of sight at moderate inclination, is typically partially 
ionized and warm, corresponding to typical warm absorber conditions.  Gas close to the axis, which is 
traversed by lines of sight at low inclination, is fully ionized 
and close to the Compton temperature, $T_{IC}\sim 10^7$K.  Lines of sight to the center which pass through 
the throat never have Thomson depth greater than unity. Our results neglect the  $\leq 10\%$ contribution 
of the multiply-forward scattered  radiation at energies $\geq$ 6 keV. 

The source function is calculated separately from the evaluation of equation \ref{eq1} 
and is stored on a spherical grid centered on the X-ray source.   Equation \ref{eq2} is 
evaluated by stepping outward from the source in radius.  The source 
spectrum is assumed to be a $\gamma$=2 power law, where $\gamma$ is the spectral index in photon number.  
For each direction we integrate the optical depth outward.
The opacity $\kappa$ at each position depends on the ionization, excitation and temperature via the 
local density and the radiation field.  We do not calculate $\kappa$ directly but rather use a stored, precalculated
tabulation in which the opacity is assumed to be a function of the ionization parameter, $\xi=4 \pi F/n$, where $F$
is the total ionizing flux locally, and $n$ is the gas density.  We emphasize that at each point $xi$ is calculated 
using the local attenuated spectrum.  However, the tabulation is calculated using an 
illuminating spectrum  which is the same as the unattenuated spectrum from the source, and so 
it is not fully self-consistent when implemented in our calculation, in the sense that it does not employ the 
same spectral shape as is found at each point.  
This is necessitated by computational limitiations:  it is prohibitive to calculate the ionization balance, temperature, opacity and emissivity separately for each spatial zone in our models.  Practicality 
dictates that we use a saved precalculated table of these quantities which are scaled according to the 
local ionization parameter.  Our procedure does employ the appropriate value of the ionization 
parameter at each point,  and it does have the appropriate behavior in regions where the central 
radiation is weak due to attenuation.   Thus the ionization etc. at each point are calculated using a radiation 
field 
whose {\it shape} corresponds to the unattenuated radiation field but whose {\it strength} corresponds 
to the attenuated radiation field.    At each radius along each radial vector, 
we store the source function $S(\varepsilon, {\bf r})$ and the opacity $\kappa(\varepsilon,{\bf r})$ for 
later use in equation \ref{eq1}.  We adopt 
a grid consisting of 128 radius zones and 20 x 20 angle zones for the calculation and storing of the source 
function.

Our dynamical models are calculated using a cylindrical coordinate system with the central axis of the torus along 
the z direction.  The dynamical equations are calculated explicitly only for azimuthal angles 0 to $\pi$; 
the other half space is obtained by replicating this.  Typical resolutions for the dynamical calculations 
are 200x20x200 in cylindrical radius, azimuthal angle (0 to $\pi$) and z.  For the purpose of calculating 
the spectra we interpolate the density and velocity field onto a uniform cartesian grid in three dimensions with 128
zones in each dimension.  
The formal solution evaluation, equation \ref{eq1}, uses a cylindrical coordinate system with the X-ray 
source at the origin and the viewing direction along the axis.  
For the calculation of both the source function and the spectrum we set the density 
in the two spatial zones nearest to the center to zero.


\section{Results}\label{results}

The warm absorber spectrum can be influenced by various factors including:  the dominant forces affecting the gas dynamics, the total 
amount of gas in the region, the time elapsed in the dynamical simulation, the viewing angle relative to the rotation axis, and the 
luminosity of the central source. 

A simple assumption is that we view AGNs is a steady state, i.e. that 
the accretion flow conserves mass across the spatial region of interest and 
over timescales  long compared with the time it takes for a parcel of gas 
to traverse the region.  Our models, in contrast, calculate the dynamics of gas which is introduced to the 
computational domain at the beginning of the calculation, and then is allowed to evolve according to the 
predictions of the MHD/Hydro equations.  No inflow or resupply of gas into our computational domain is assumed.   Eventually, most of the gas in our models leaves the computational domain,
either via accretion or via outflow in a warm absorber wind.  In order to simulate a steady-state AGN, we choose
to analyze our dynamical models at a time which is chosen so that the model appears to be evolving slowly, i.e. 
many dynamical times after the start of the simulation.    But we do not consider times which are late enough
that most of the torus gas has been lost.  We choose to show results at  60 dynamical times after the 
start of the simulation, which satisfies these conditions.

The ionization and heating of the torus, and the warm absorber flow, depend on the intensity of the 
illuminating radiation field.  The total intensity is specified by the Eddington ratio, i.e. the ratio 
of the total bolometric luminosity assumed from the central source to the Eddington value for a $10^7 M_\odot$ 
black hole.  The ionization balance and heating depend on the radiation field in the energy band above 
13.6 eV, and it is conventional to use the integrated luminosity from 1 to 1000 Ry $L$ in defining the 
ionization parameter $\xi=L/(nR^2)$.  The shape of the spectrum affects these definitions and their relationship;
here we choose an extremely simple spectrum which is a single power law with photon number index $\gamma$=2.
In this case, the total bolometric luminosity and the ionizing luminosity are related by the factor 
${\rm ln}(13.6 eV/\varepsilon_{min})$ where $\varepsilon_{min}$ is the minumum energy used in the calculation 
of the bolometric luminosity.
More realistic spectra may have a flatter slope at energies above $\sim$1 keV, and steeper slope at lower 
energies.  But we are confident that these differences will not change the qualitative results that we find here.

We consider six dynamical models.  These are summarized in table 1.  They span the different initial field 
geometries described in section \ref{models}, and include different values of the central source luminosity as 
described by the Eddington ratio.  All are analyzed 
at various viewing angles.

\begin{deluxetable}{crrrrrrrrrr}
\tabletypesize{\scriptsize}
\tablecaption{model parameters \label{modelparameters}}
\tablewidth{0pt}
\tablehead{
\colhead{Model}&\colhead{L (erg s$^{-1}$)}&\colhead{M$_{Torus}$ (gm)}&\colhead{note}}
\startdata
1 & 3E+45 & 1.45E+38 & TOR \\
2 & 1E+45 & 1.44E+38 & TOR \\
3 & 5E+45 & 1.45E+38 & TOR \\
4 & 1E+45 & 9.15E+36 & SOL\\
5 & 3E+45 & 8.90E+36 & SOL \\
6 & 3E+45 & 1.06E+37 & hydro \\
\enddata
\end{deluxetable}

\begin{figure*}[p]
\includegraphics*[angle=0, scale=0.35]{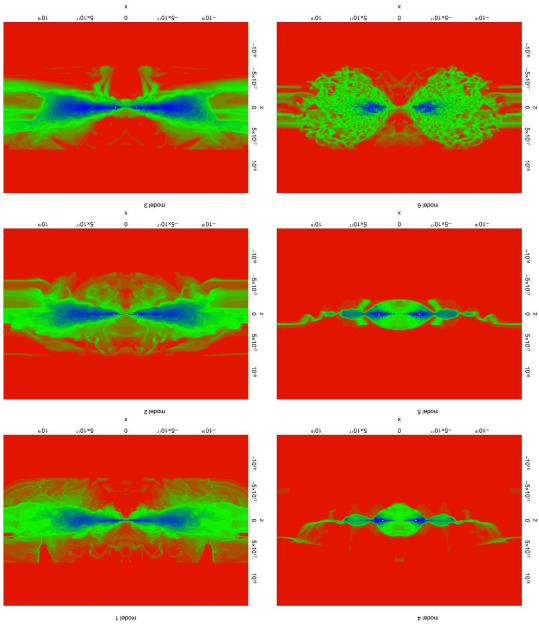}
\caption{\label{fig1} Density in a vertical plane at 60 dynamical times for the models used in this work.  Colors correspond to log of density  as shown in color bars. 
Axes correspond to position relative to the central black hole; each panel 
spans 3 $\times 10^{18}$ cm,  
the horizontal and vertical directions.  Models are derived from 3D models; shown are slices at a 
fiducial azimuthal angle. Model numbers correspond to table \ref{modelparameters}.}
\end{figure*}

The density distributions in a vertical plane in our models, at a time of 60 dynamical times, are shown in figure 
\ref{fig1}.  It is apparent that the initial magnetic field strongly influences the density distribution.  The SOL models 
have density distributions which are more confined in the vertical direction than the TOR models.  In addition, as 
shown in \citet{DoraKall17}, they evolve more slowly with time than the TOR models.  
This is due to the stabilizing role of the poloidal magnetic field in the SOL models.  
The hydro models have the magnetic forces turned off.  They evolve more rapidly than either of the MHD models
 and generally have a density distribution is more extended in the vertical direction.

All of our models have column densities in the horizontal plane which are $\geq 10^{25}$ cm$^{-2}$, making them 
Thomson-thick when viewed at high inclination.  The column densities along the z axis are much smaller, though 
the spatial resolution of our grid used for the radiation transfer calculations often cannot resolve regions with 
column density per spatial zone is less than $\sim 10^{22}$ cm$^{-2}$.

\subsection{The Ionization Parameter Distribution}

\begin{figure*}[p]
\includegraphics*[angle=0, scale=0.6]{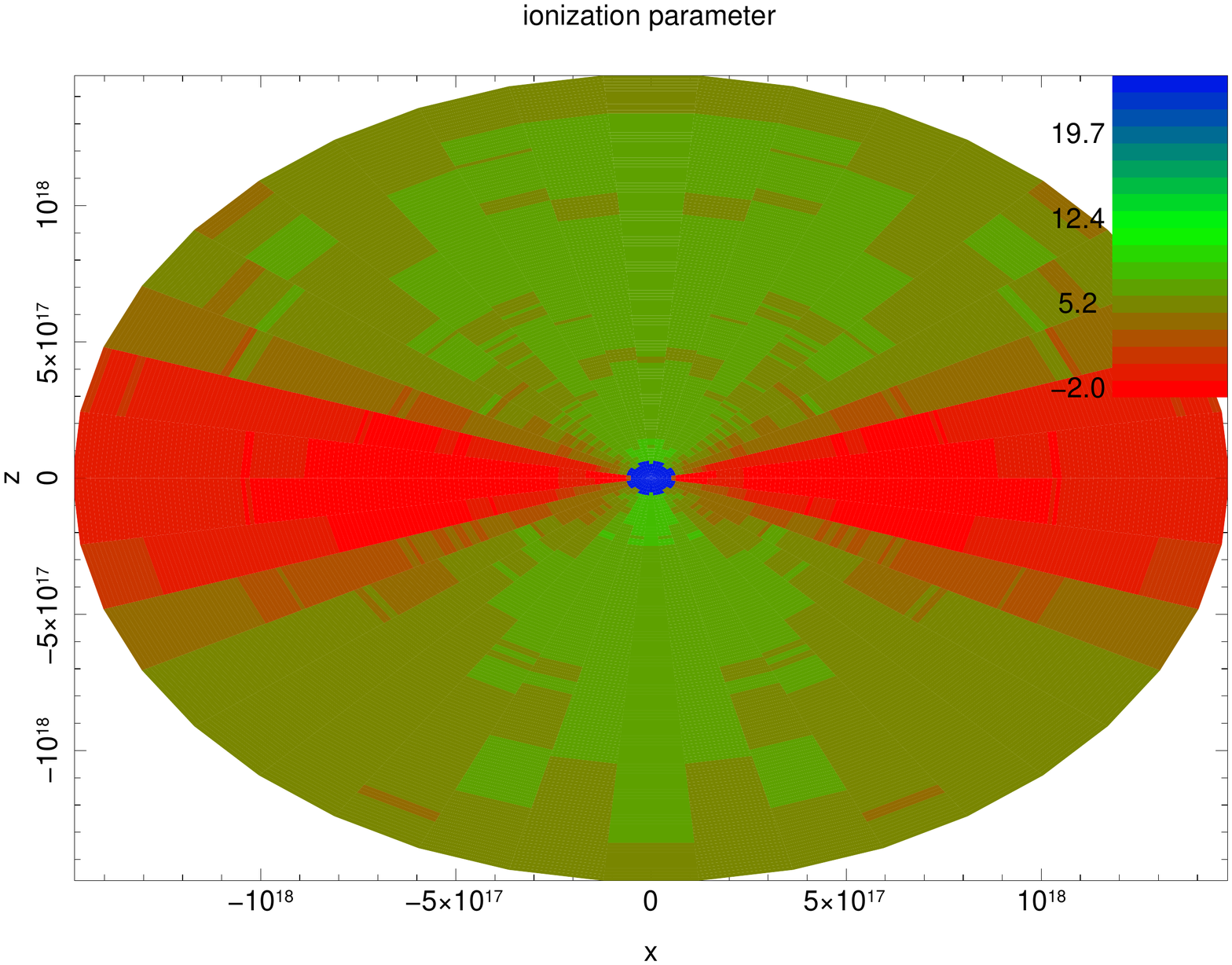}
\caption{\label{figxiplot} Ionization parameter contours vs. position for model 1 }
\end{figure*}

Ionization parameter is defined as $\xi=4 \pi F/n$.  $F$  is the local flux 
integrated over energy from 1 - 1000 Ry.  Values of this quantity ${\rm log}(\xi) \leq 0$, approximately, 
are low ionization or neutral.  If so, the opacity of the gas approaches the opacity of neutral 
material (eg. \citet{morr83}) and the effective absorption cross section per hydrogen atom is
crudely $\sigma_{Eff}\sim 2.5 \times 10^{-22} \varepsilon_{keV}^{-3}$ cm$^2$.  For larger values of $\xi$ 
the gas becomes more ionized and the effective absorption cross section approaches the Thomson
cross section for ${\rm log}(\xi) \geq 4$.   Figure \ref{figxiplot} shows contours 
of ${\rm log}(\xi)$ vs. position for model 1.   This demonstrates that the gas in the orbital 
plane remains  neutral and negligible directly transmitted radiation escapes in that direction.  In the direction perpendicular 
to the orbital plane the gas is ionized and the total optical depth is lower.

\subsection{The absorption measure distribution (AMD)}

Figure \ref{figxiplot} demonstrates that the absorption spectrum seen by a distant 
observer is a superposition of absorption spectra from gas at a range of ionization 
parameters.  A convenient way to describe this is the absorption 
measure distribution or AMD \citep{Holc07}.  This is the distribution of column 
densities per unit ionization parameter vs ionization parameter along the line of sight.  
\citet{Holc07} define it as ${\rm AMD}=dN/d{\rm log}(\xi)$.  We give our results in 
the slightly modified form: ${\rm log(AMD)}=d{\rm log(N)}/d{\rm log}(\xi)$.
This quantity, as calculated by \citet{Holc07} and others,  is derived from the observed spectra
by fitting one or more lines from individual ions in order to derive a column density for 
each ion.  The ion column densities can then be expressed as a sum over a distribution of 
column densities vs. ionization parameter, $N(\xi)$ by using  photoionization models 
for the charge state distribution as a function of ionization parameter. 

In our calculations, we do not need to derive the ionization parameter distribution from 
fitting individual ion spectra.  Our MHD/Hydro models provide the gas density as output, and we can calculate the 
ionization parameter for each spatial zone in our models directly.  This quantity includes the fact that the 
flux coming from the center at any given point is subject to attenuation as it traverses the gas; this is 
better described in section \ref{models} where we describe the source function calculation.  Similarly, we know the 
column density associated with each spatial zone.  So we can calculate the AMD along a given line of sight 
by simply summing these quantities along the line of sight. 
During our integration of equation \ref{eq1} for each spatial zone along the line of sight, which we can call zone $j$,
 we calculate the ionization parameter for that zone as $\xi_j= 4 \pi F_j/n_j$ where $F_j$ is the flux from the center 
in zone $j$, and $n_j$ is the gas number density in zone $j$.  
The column density in zone $j$ is $N_j=n_j \Delta R_j$ where $\Delta R_j$ is the path length of the 
line of sight through zone $j$.   
We set up a discrete grid of  ionization parameter, ${\rm log}(\xi_i$) with bin size 
$\Delta{\rm log}(\xi_i)=0.5$, and then we step through the values of ${\rm log}(\xi_i$), and 
for each one we find the zone column densities $N_j$ with ionization parameter values in that range, and sum them, i.e.
$AMD_i=\Sigma_j N_j \vert_{\xi_i-0.25 \leq \xi_j \leq \xi_i+0.25}/\Delta{\rm log}(\xi_i)$.
The column density is the 'equivalent hydrogen' column density, i.e. the column density of hydrogen, 
both neutral and ionized.  This quantity
describes the relative quantity of gas at each 
ionization parameter, as it affects the absorption spectrum along a given line of sight.  
The total column density along that line of sight is then   
$N_{tot}=\Sigma_i{AMD}_i \Delta{\rm log}(\xi_i)$.  
Clearly, the AMD depends on all the input parameters describing the model, including the
line of sight chosen.

\begin{figure*}[p]
\includegraphics*[angle=270, scale=0.6]{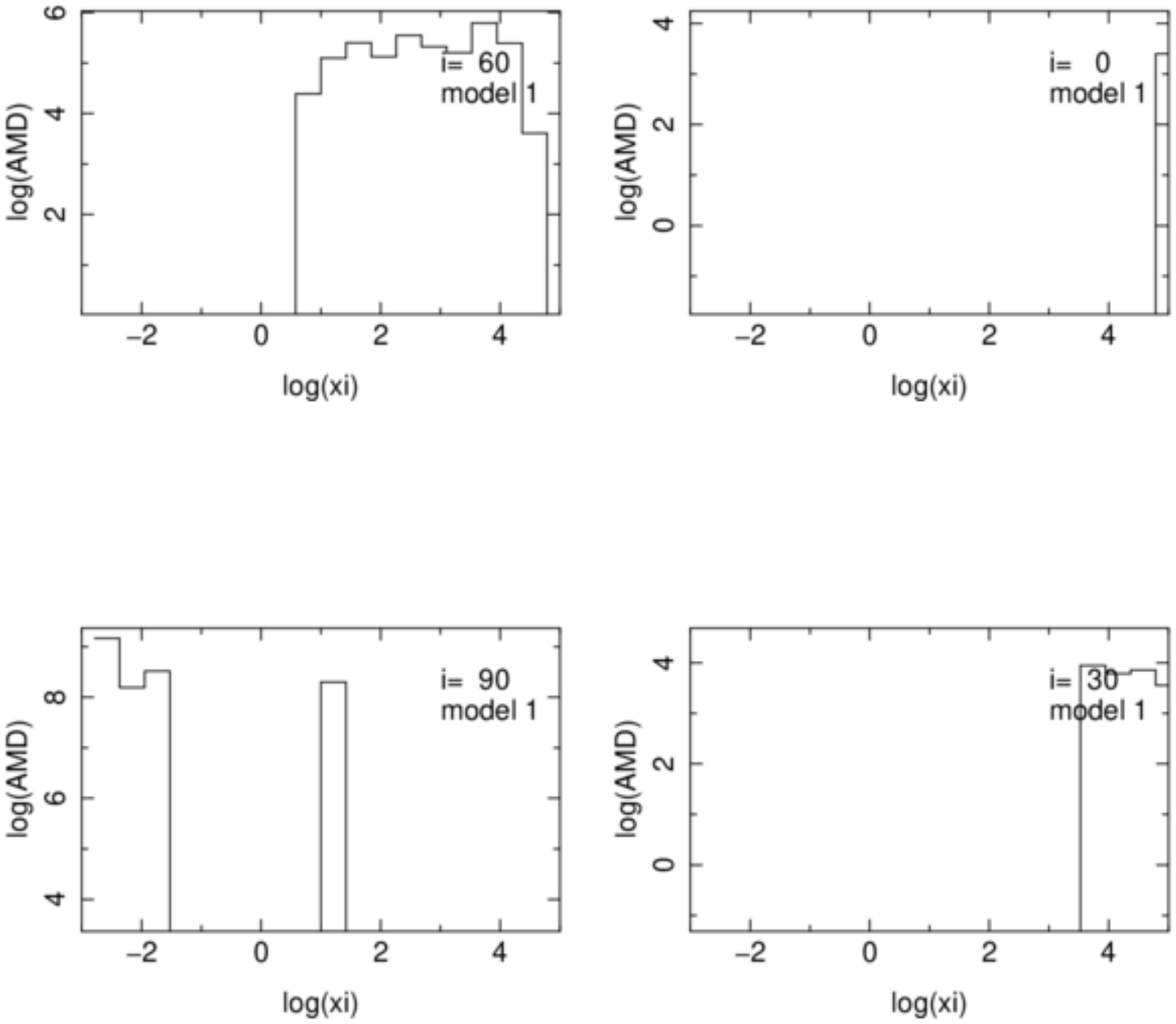}
\caption{\label{figamd1} Absorption Measure Distribution (AMD) vs inclination.  AMD is plotted 
as log in units of $10^{18}$.}
\end{figure*}

Sample results for AMD are shown in figure \ref{figamd1}, for model 1 as a function of the 
inclination angle.  Here and in what follows we define inclination $i$ as the angle between 
the viewer's line of sight and the axis of the torus (perpendicular to the orbital plane).  
This figure plots the log of the AMD in units of $10^{18}$ cm$^{-2}$.
This shows that the distribution changes from all highly ionized (log($\xi) \geq$ 4) for $i$=0
to nearly neutral for $i$=90$^o$, and that there is a broad distribution 
of partially ionized material at intermediate inclinations.  

\begin{figure*}[p]
\includegraphics*[angle=0, scale=0.5]{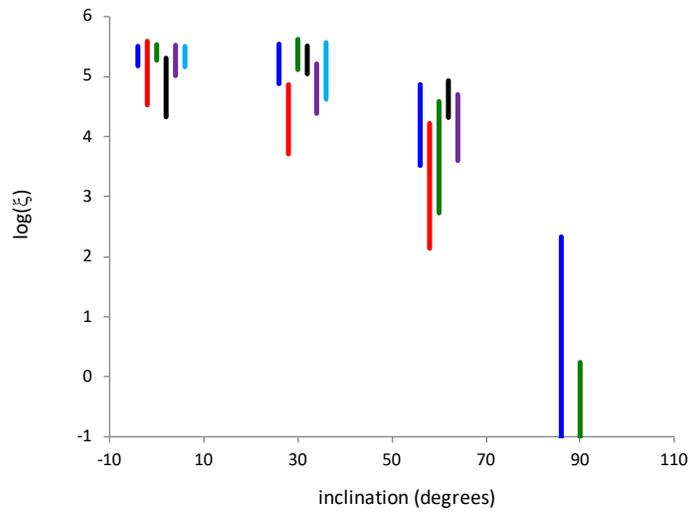}
\caption{\label{fig10} Mean and dispersion values of the ionization parameter derived 
from the AMD distribution are plotted for our 6 models vs. inclination.  The colored bars 
correspond to the models numbered in order of increasing index:  dark blue: model 1,
red: model 2, green: model 3, black: model 4, purple: model 5, light blue: model 6}
\end{figure*}

The AMD can be described in a simpler way by calculating the mean and the dispersion of the distribution for
each model and viewing angle.  These quantities are shown for our models in figure \ref{fig10}.  
For many of the models, notably the ones at lower inclination, there is significant AMD at ionization 
parameters ${\rm log}(\xi) \geq$4.  This leads to larger values of the mean ionization parameter than would be 
inferred from results in figure \ref{figamd1}.
This shows that all the models have similar behavior.  However the inclination angles where intermediate ionization 
gas appears differs:  of the TOR models, the low luminosity model (2) shows a slightly greater range of 
inclinations where intermediate ionizations appear.  The SOL models (4 and 5) show very few or no inclinations
where intermediate ionization parameters appear.  We attribute this to the compact distribution in the vertical direction of these 
models, and the fact that they evolve little during the time we have computed.  Model 6, the pure hydro model,
is the most extended in the vertical direction.  It shows high ionization gas at low inclination angles 
($i \leq 30$) and at higher inclinations there is negligible flux transmitted by the torus.

\begin{figure*}[p]
\includegraphics*[angle=0, scale=0.5]{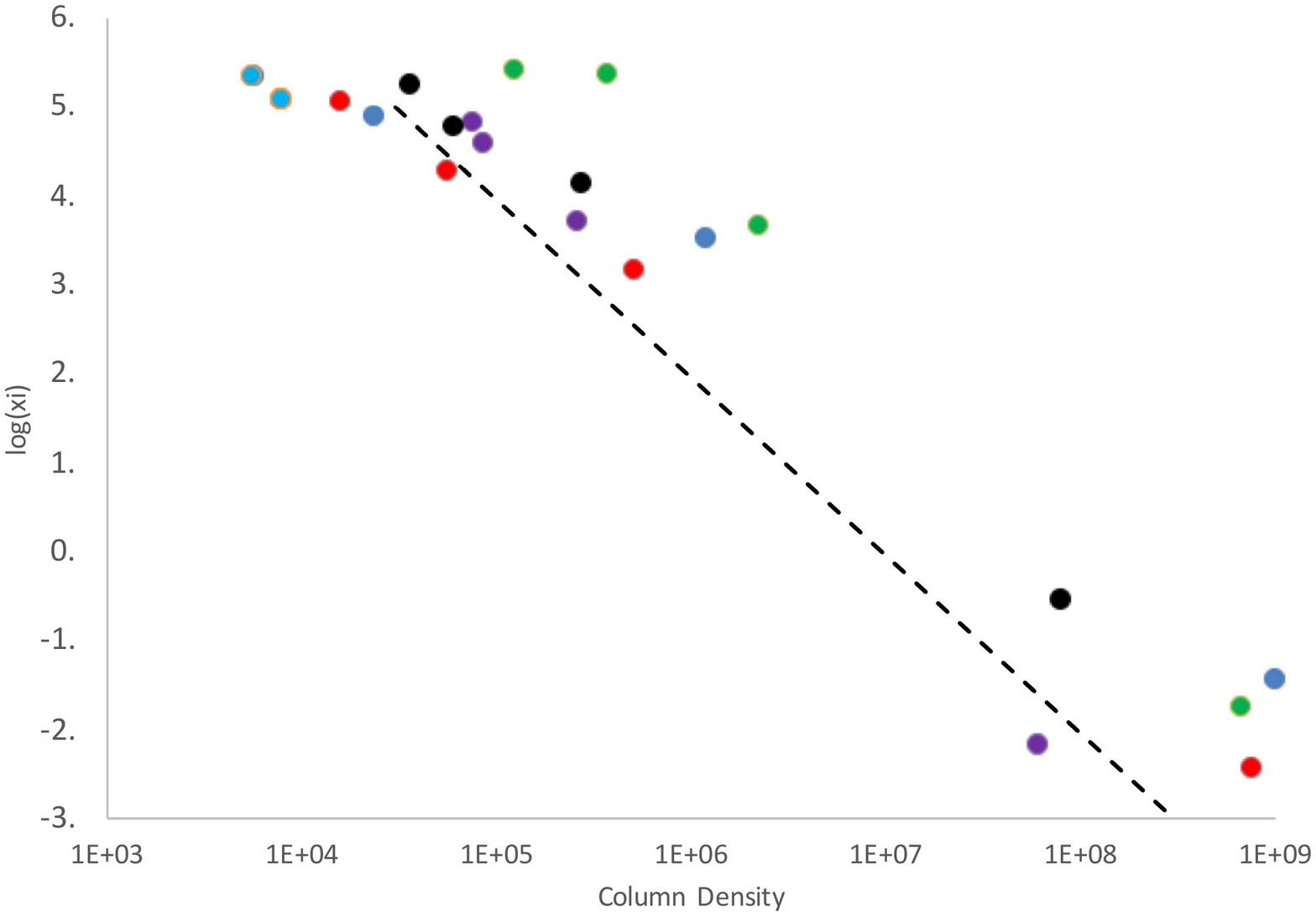}
\caption{\label{fig11} Mean  values of the ionization parameter derived 
from the AMD distribution are plotted for our 6 models vs. the column density along the line of sight (in units of 10$^{18}$ cm$^{-2}$) for each model.  
The colored points 
correspond to the models numbered in order of increasing index and are the same as in figure \ref{fig10}:  dark blue: model 1,
red: model 2, green: model 3, black: model 4, purple: model 5, light blue: model 6}
\end{figure*}

This can also be viewed in the $\xi$ vs. column plane, as shown in figure \ref{fig11}.  This shows that the models 
approximately span a line from high $\xi$ and low column to low $\xi$ and high column density.  They approximately
follow a line corresponding to $\xi \propto N^{-2}$, shown as the dashed line in this figure.  This corresponds
to the expected scaling for a constant density sphere of gas viewed at different radii, i.e. if 
$\xi \propto R^{-2}$ and $N \propto R$, where $R$ is the distance from the center.  The region corresponding 
to observed warm absorbers lies at intermediate $\xi$ and $N$; models differ according to the range of $\xi$ 
at intermediate $N$ (i.e. the AMD), and also the extent to which they follow the simple scaling.

We emphasize that this scaling is an approximate, phenomenological pattern which we find in our model results.
The analogy with a constant density sphere is not intended to be predictive or accurate for quantitative 
model fitting.   Spectra and AMDs from individual lines of sight show a rather different behavior:  that the 
AMD is approximately flat as a function of ionization parameter (eg. figure 3).  This can be compared with 
observationally derived AMDs \citep{beha09,Holc07,Holc10} which often show a trend of AMD which increases with 
ionization parameter.



\subsection{Differential Emission Measure}

\begin{figure*}[p]
\includegraphics*[angle=90, scale=0.8]{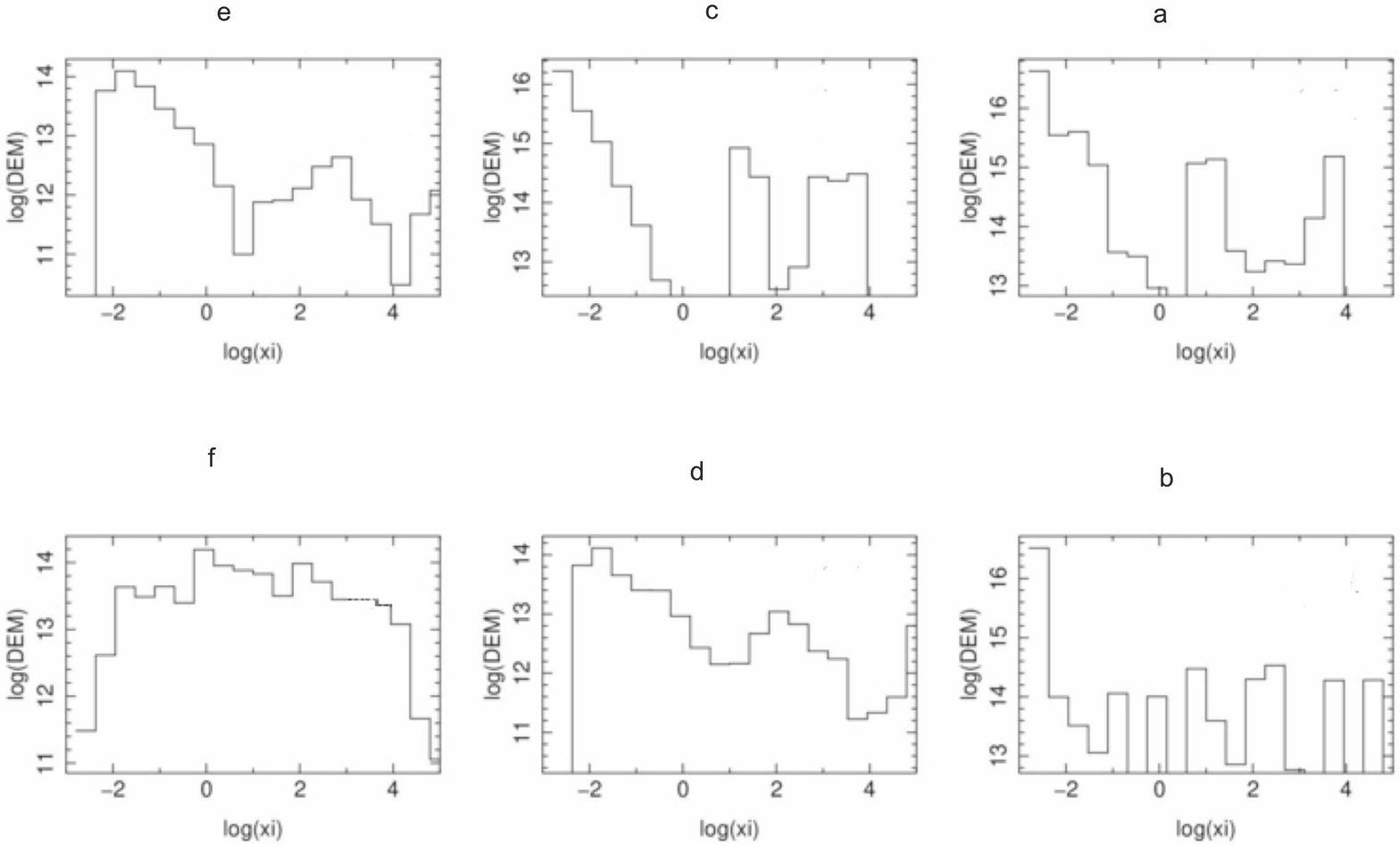}
\caption{\label{figdem1} Differential emission measure distributions for our models, as viewed at 
$i$=60$^o$, in units of 10$^{54}$ cm$^3$.  Panels labels correspond to model numbers: a=1, b=2, c=3, d=4, e=5, f=6.}
\end{figure*}

The emission spectrum is affected by the differential emission measure (DEM).
This quantity is defined as $DEM={\rm d}(EM)/{\rm dlog}\xi$ where 
$EM=\int{n(\xi)^2 dV}$.
This quantity is only weakly dependent on the viewing angle.  
It is plotted in figures \ref{figdem1} for viewing inclination $i=60^o$.
The DEMs look qualitatively similar, though the range of ionization parameters 
is greater for the high luminosity model 3 and lowest for model 2.  

\subsection{Emission vs. Absorption}

\begin{figure*}[p]
\includegraphics*[angle=0, scale=0.5]{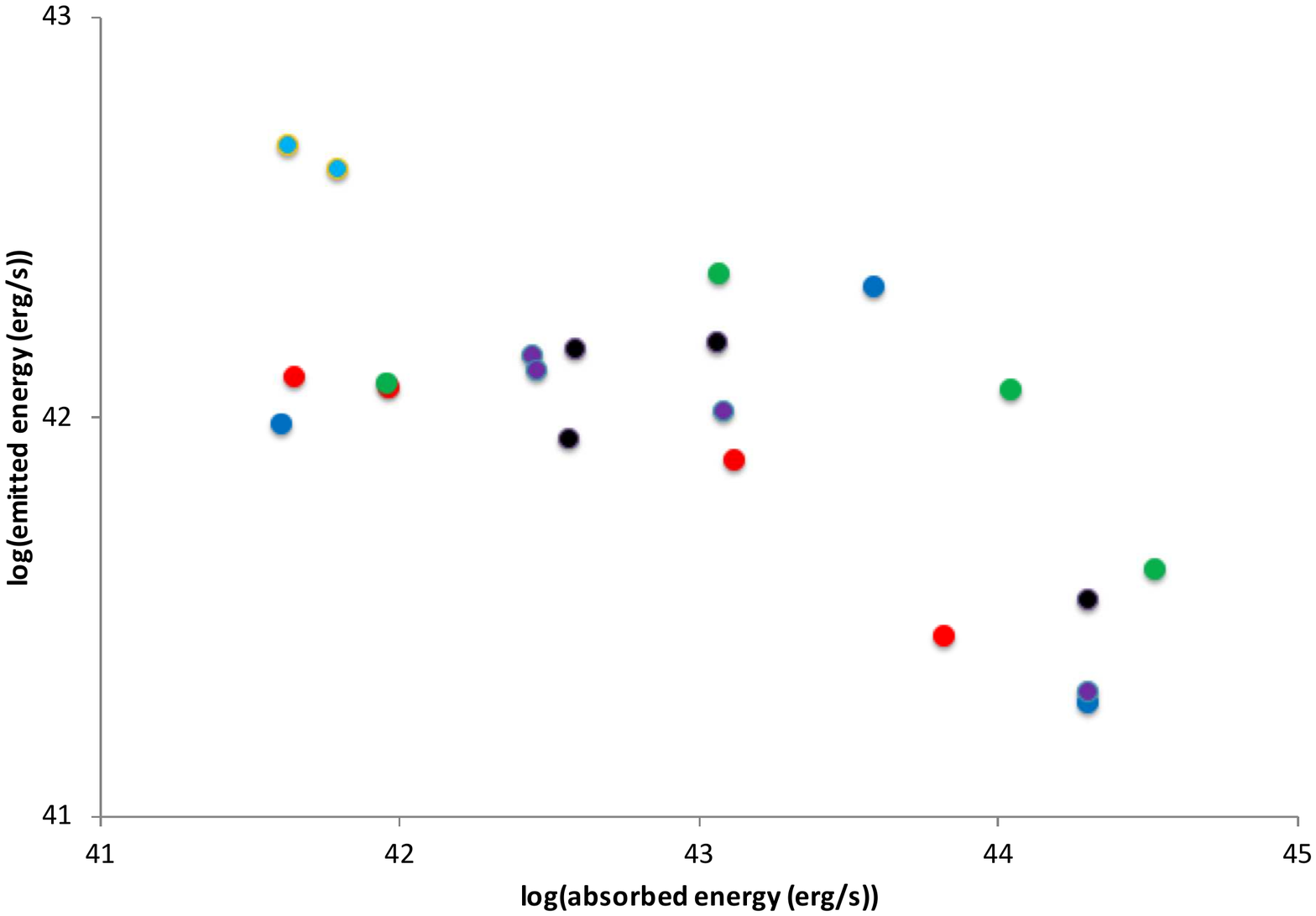}
\caption{\label{figabsem} The relative amounts of absorbed vs. emitted energy for our models. Colors are
the same as in figures \ref{fig10} and \ref{fig11}}
\end{figure*}

The absorbed and emitted energies vs inclination are shown in figure \ref{figabsem}.  This shows that the absorbed fraction generally increases with inclination, while the emitted fraction is more nearly isotropic.  At the highest luminosity, model 3 shows that the absorbed fraction is only weakly dependent on inclination.  However, our results show that the ionization of the absorbing material does depend on inclination, cf. fig \ref{fig10}.

\subsection{Fitting to Observed Warm Absorbers}


A key check on warm absorber spectra is direct fitting against observed warm absorber data.
The dataset we compare with is the 900 ksec HETG spectrum of NGC 3783 \citep{Kasp01}.  Model fits to this 
 spectrum have been published by \citet{kron03,chel05,fuku18,gonc06,goos16,mao19}.  As shown by \citet{goos16}, this spectrum 
fits to a phenomenological model consisting of a broad AMD distribution between ${\rm log}\xi$ = -1 and  ${\rm log}(\xi)$ = +0.5 and also between
log($\xi$) = 2 and ${\rm log}(\xi)$ = 4.  There is a gap between ${\rm log}(\xi)$=1 and ${\rm log}(\xi)$=2; a significant amount of 
gas in this range will produce absorption in the iron m shell UTA near 15 $\AA$ which is not observed.  This gap is interpreted
as evidence for thermal instability, i.e. that photoionization equilibrium gas in this range will cool or heat 
rapidly \citep{adhi15, goos16, roza06, dyda17}.  This explanation requires that, if the different ionization parameters coexist in approximately the same spatial region, then the gas density must change in response to X-ray heating.  The dynamics must allow this to occur.  An example of such a situation is a gas in pressure equilibrium.  If so, the sound crossing time for the region must be shorter than other relevant timescales, such 
as the flow timescale through the region.

We have taken the output from our spectral calculations and used them to directly fit to observed $Chandra$ HETG warm 
absorber spectra.
We adopt a continuum consisting of a power law plus a low energy thermal blackbody component with kT=0.1 keV.  
We include a simple cold absorption model ('wabs' \citet{morr83}) to account for the interstellar medium.  
We convert our warm absorber spectra into a multiplicative 'analytic model' for use in the {\sc xspec} analysis program 
\citep{arna96}.  That is, in {\sc xspec}, 
we use the model command `model wabs*ourmodel*(bbody+pow)', where 'ourmodel' is our analytic model created from the spectra
calculated using equation \ref{eq1}. The 'wabs'
model component has only one free parameter, the column density, which corresponds to the equivalent hydrogen 
column density of a neutral gas with cosmic abundances.

The 'ourmodel' model has effectively three free parameters.  
One is the model index, corresponding to the model number (column 1) in table 1.  
Clearly, each of the models in table 1 embody 
different assumptions about the magnetic field and the luminosity of the central source, so these parameters 
are implicit in the model index.  The second  free parameter is 
the inclination; for each value of the model index we calculate 4 or 5 inclinations, including 0, 30, 60, and 90 degrees.
We emphasize that we do not 'fit' for the values of these two free parameters; we simply step through them and examine 
the fit results for each.  The third free parameter is a redshift which is applied to the model. 
Of course, our models for the spectra also include the Doppler shifts associated with the flow in our MHD/Hydro models; the 
redshift is an additional wavelength shift which may or may not be needed to fit the observed spectrum.
This redshift accounts for the possibility that the gas density distribution will produce a good fit to the observations, 
but the flow speeds in the MHD/Hydro models differ from those observed. So, for example, in our fits to NGC 3783 below, 
the redshift of the galaxy is z=0.0085.  If our dynamical models are correct in both the shape of the absorption and in 
its wavelength scale, then we would expect the best fit to occur for this value of redshift.  
 We emphasize that the strength of all absorption features and the overall shape of the warm absorber 
absorption are hardwired into 'ourmodel' for given values of the model index and the inclination; 
we do not vary any quantities 
affecting the depth or shape of the warm absorber absorption during the fitting process.  
Our procedure is to step through all the model synthetic spectra, both the model number 
and the inclination.  For each of these we allow the wabs column density, redshift of the warm absorber, 
the normalization and slope of the power law, and the normalization of the blackbody
to vary in order to get a best fit.

\begin{figure*}[p]
\includegraphics*[angle=90, scale=0.6]{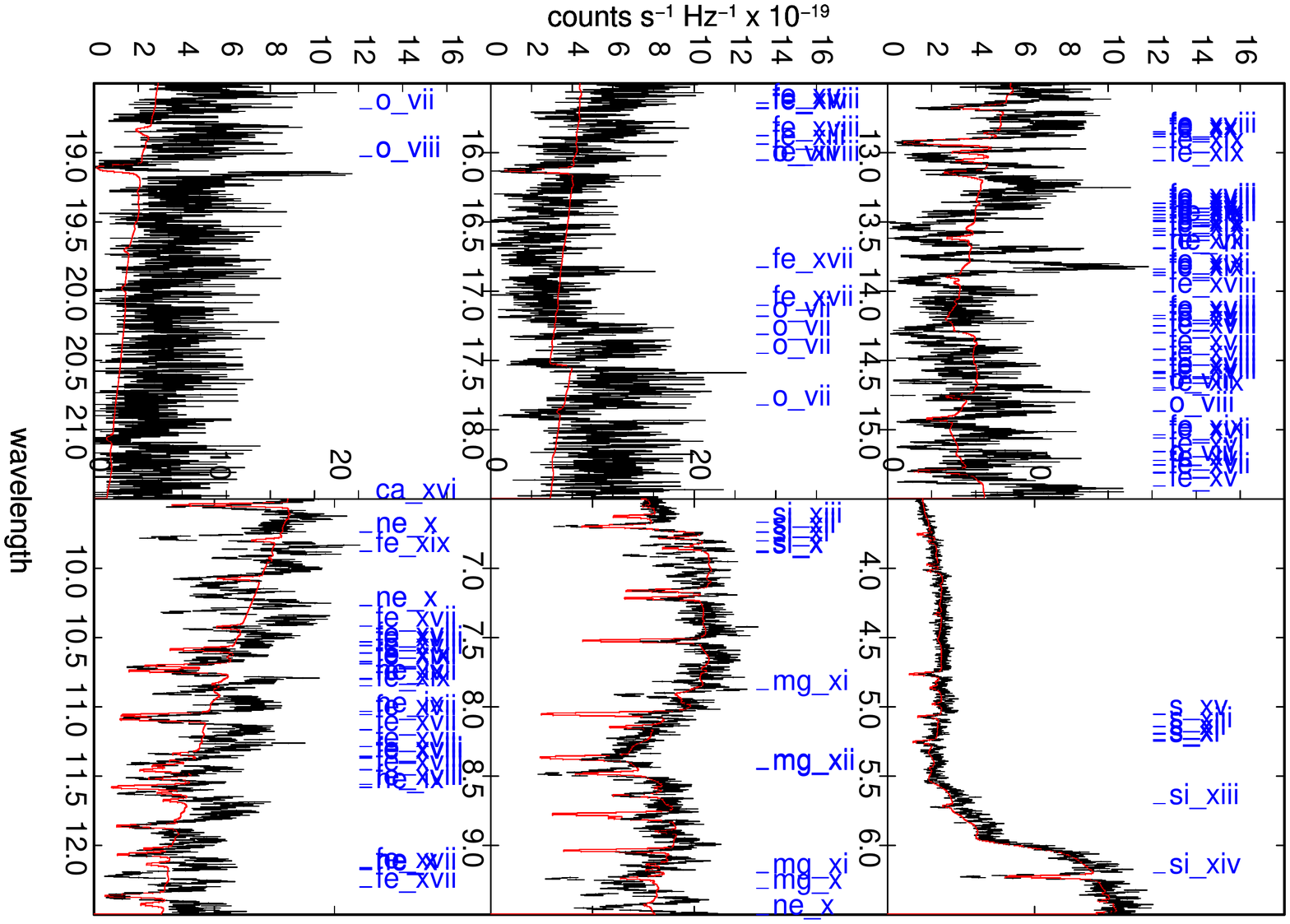}
\caption{\label{fig12} Fit to the Chandra HETG spectrum of NGC 3783 using our model 3 viewed at $i=60^o$.  Strong lines are 
labeled.}
\end{figure*}

  With this procedure 
we obtain $\chi^2/\nu$ values for each of our models, i.e. models 1 - 6 at inclination angles between 
0 and 90 degrees.    The best fit value is  $\chi^2/\nu$=2.3. 
The results of one such fit is shown in figure \ref{fig12}, using a model spectrum from model 3 viewed at $i=30^o$.
This shows that the model reproduces most strong absorption features from the high ionization gas, i.e. at wavelengths 
less than $\sim 15 \AA$.  Features from gas with ionization parameter $\xi\leq 0$ are absent or weaker than observed, such 
as the iron m shell UTA at 16 -- 18 $\AA$.  The line widths in the model are comparable to the observed widths.  This model 
gives  $\chi^2/\nu=2.311$  This can be compared with a fit to a phenomenological model constructed using {\sc warmabs} 
\footnote{https://heasarc.nasa.gov/lheasoft/xstar/xstar.html} which gives $\chi^2/\nu=1.62$ using three ionization parameter 
components.

The redshift of NGC 3783 is 0.0085 \footnote{http://simbad.u-strasbg.fr/simbad/}.
Our best fitting model,  model 3, requires a redshift 0.0096.  By comparison, the best 
fit to  {\sc warmabs} warmabs phenomenological model has redshift z=0.007, 
corresponding to an outflow velocity of 450 km s$^{-1}$.  
Thus, the gas in our model 3 is predominantly outflowing with a speed which is  $\simeq$ 330 km s$^{-1}$ greater 
than is indicated by the observations. 
Other models produce outflow speeds which are closer to the observations:  with model 1 we obtain fits with statistical quality comparable to that shown in figure \ref{fig12} using 
redshift z=0.0085.  

\subsection{Emission Spectrum Fitting}

We have also attempted to fit our models against observed emission spectra.  We adopt NGC1068, in which it is 
thought that the direct line of sight to the center is obscured, so that the observed spectrum is dominated 
by scattering and reprocessing in the torus.  We compare the Chandra HETG spectrum \citep{kall14} 
against our models observed at inclination $i$=90$^o$.
We use the same fitting procedure as for absorption, except we treat the model as an 'additive' model 
rather than a multiplicative model.  The continuum is included as part of our models.  We allow the wabs 
column density to vary in order to get a best fit.  The free parameters associated with the emission 
component derived from our models, are the redshift and the normalization.  

\begin{figure*}[p]
\includegraphics*[angle=90, scale=0.6]{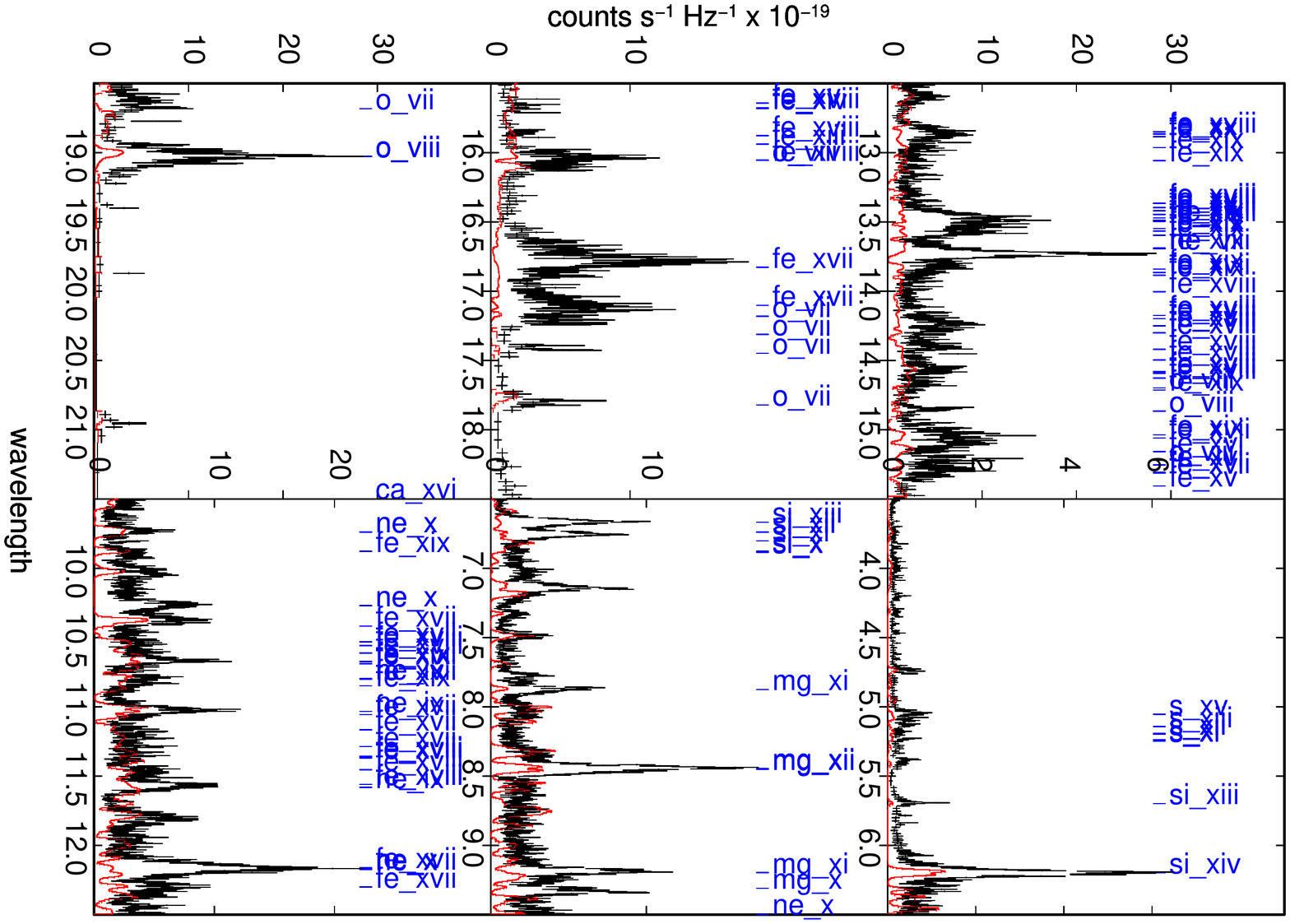}
\caption{\label{fig13} Comparison of model 6 synthetic spectrum observed at inclination $i=90^o$ with the 
observed Chandra HETG observation of NGC1068.  Strong lines are labeled.}
\end{figure*}

A sample fit is shown in Figure \ref{fig13}.  This shows that, although our models produce strong emission and 
many of the observed lines, the overall agreement is poor.  In particular, 
the model produces strong emission in the region of the Fe L lines, 
10-12 $\AA$, but does not produce sufficient emission in the Si K lines near 6 $\AA$.
This is consistent with the fact that the differential
emission measure from our models is not as broad as that for NGC1068 (see eg. \citet{kall14} fig 11).  
Also, the phenomenological fits required enhanced abundances for some elements 
(O, Ne, Ar, Ca); here we employ cosmic abundances.

\section{Discussion}

We have calculated MHD/Hydro models for obscuring gas in Seyfert galaxies, for various 
assumptions about the luminosity of the central source and the initial field configuration.  Using the output
from these we have calculated synthetic spectra using a formal solution of the equation of radiation transfer
as a function of viewing angle relative to the central axis.  The results of these simulations show that all 
the models have regions with conditions which very crudely might be expected to produce warm absorber spectra.
That is, for viewing angles $\sim$ 30 -- 60 degrees they view through material with ionization parameter 
comparable to the dominant values observed.  However, when considered at a  more detailed level, 
several do not contain sufficient column density at these ionization parameters to produce the strong 
absorption seen from well known astrophysical sources.  These include the models employing the SOL initial 
field geometry, i.e. models with enhanced poloidal magnetic field, 
which tends to produce the most stable, compact, and long-lived torus configurations.   
Conversely, models in which the dispersal or evaporation of the torus is most rapid tend to produce 
warm absorber spectral closest to those observed.  These models include the TOR initial magnetic field 
geometry, i.e. models with predominantly toroidal magnetic field,
 at high Eddington ratio, and also the pure hydrodynamic models.  The ability to produce realistic 
warm absorbers is clearly connected to the absorption measure distribution (AMD), and the most realistic 
models have the broadest AMD.  Thus, in the context of our models, the most realistic models are those in which 
the torus is most dispersed by X-ray heating and unbalanced dynamical forces.

Our models differ from many previous models for warm absorbers in that we begin with dynamical models 
for gas at 1 pc near a black hole and calculate the spectrum from that.  The free parameters are the 
viewing angle, the 
central source luminosity or Eddington ratio, the geometry or strength of the initial magnetic field, the shape 
of the ionizing spectrum, and the masses of the central source and the initial gas in the computational domain.
Of these, we only consider 6 combinations of parameter values for the Eddington ratio and field geometry, 
as described in table 1.  All our models have the same central source mass, ionizing spectrum and initial mass in 
the computation domain.  We consider 4 viewing angles or inclinations for each of the MHD/Hydro models.
When fitting to observed spectra, we also allow the redshift to vary.  This is in contrast with most 
other such efforts, in which various simple template models are used to fit to the observations.  The parameters
describing these templates include the ion column densities \citep{Kink02, Kasp02}, or the ionization parameter
and total column density \citep{Holc10}.  These quantities can be compared with the results of dynamical models,
but they do not uniquely test specific dynamics.  Work which is more similar to ours is that of 
\citet{fuku18}, who use the results of models for magneto-rotational winds to specify the kinematic 
properties of warm absorber gas.  These models have as a free parameter effectively a fiducial density 
which can be varied when constructing spectra for comparison with data.  Nonetheless, our models remain the only ones to our knowledge in which the dynamical results completely determine the observed spectrum without any additional free parameters.

None of the models considered here exhibits strong evidence for thermal instability
in the AMD.  We attribute this to the fact that the gas speed all the models is supersonic, and there is not
sufficient time for pressure equilibration in the flow.   This is a key constraint on any model for the
dynamics of warm absorber gas.

In considering what dynamical models may produce warm absorber spectra which are 
closer to what is observed, it is worth considering the shortcomings of the 
models considered here.  Clearly, limitations on the numerical techniques 
can affect such simulations; our dynamical models have typical grid sizes 
of 400x100x400 in r,$\phi$,z.  This corresponds to a physical resolution element 
size $\simeq 10^{16}$ cm.  Models presented here include radiation pressure from the 
central source as a bulk force but not diffusely emitted radiation as 
it affects the equation of state of the gas.  Previous calculations by us \citep{DoraKall16} 
showed that these processes are not important in the gas which is sufficiently ionized 
to produce strong X-ray spectral features.  Models which have different magnetic field 
geometry, or a stronger external field, may be capable of producing thermal instability. 
It seems likely that such models, if they succeed, must impede the warm absorber 
outflow sufficiently to allow time for pressure equilibration locally.  This seems
in conflict with models which facilitate the outflow, eg. by having rotational 
forces drive flow along field lines.  

The SOL models presented here provide a torus which is too  compact to allow a significant 
range of inclinations which view partially ionized gas.  The pure hydro model (model 6) 
provides a density structure which is more extended than either of the MHD models.  Thus it 
will provide warm absorber spectra at smaller inclinations than the TOR MHD models (models 1, 2
and 3).  However, the pure hydro is likely artificial; the TOR model represents a plausible 
initial field configuration involving approximately equipartition field strengths.  

It is possible that, rather than being in pressure equilibrium, the low ionization gas in 
warm absorbers 
represents gas which is being transiently heated and ionized from a cold neutral state.  Such 
reservoir of cold neutral gas could be associated with dust or molecular gas moving inward from 
the AGN host galaxy.   The simulations presented here include time dependent heating 
but do not include any attempt to simulate the intermixing of such a cold component with warm gas, 
or its evaporation and associated
spectra.  The importance of this process is suggested by the coincidence 
between the location of the obscuring torus and the likely dust sublimation radius, i.e. the 
minimum radius at which dust can survive when heated by the direct flux of radiation from the 
center of the AGN.  

The results of this paper provide insight into some of the open questions 
surrounding warm absorbers.  They show that thermal evaporation from a predominantly cold optically thick 
torus can produce absorbers with ionization and optical depth similar to what is observed.  At the detailed 
level, there is a dependence on the geometry and strength of the magnetic field in the torus throat, and the 
shape of the throat opening.  Our models imply that there are many viewing angles which do not produce warm absorbers,
either because the gas along the line of sight is too highly ionized and rarefied, or because it is 
cold and opaque.  The nature of the evaporative flow is that the warm absorber gas is in a transient phase 
between its origin in the cold torus and its ultimate fate as a fully ionized outflow.  This means that there
is more unobservable fully ionized gas than we can detect with direct observations.  Our models cannot 
directly address the nature of ultrafast outflows, except to point out that such flows likely cannot coexist with 
conventional warm absobers in the 
$\sim$1 pc region, since the warm absorber, torus and associated fully ionized gas occupy almost the
entire volume.

\acknowledgements{This work was supported by NASA grant 14-ATP14-0022 thru the Astrophysics Theory Program}

%
%

\bibliographystyle{apj}

\end{document}